\documentclass[10pt,aps,prl,twocolumn,showpacs]{revtex4}

\usepackage{graphicx}
\usepackage{hyperref}

\begin{document}

\title{Environmental vs. demographic variability in two-species 
         predator-prey models}

\author{Ulrich Dobramysl}
\email{ulrich.dobramysl@vt.edu}
\author{Uwe C. T\"{a}uber}
\email{tauber@vt.edu}
\affiliation{Department of Physics, Virginia Tech, Blacksburg, 
  Virginia 24061-0435, USA}
\date{\today}


\begin{abstract}
  We investigate the competing effects and relative importance of
  intrinsic demographic and environmental variability on the
  evolutionary dynamics of a stochastic two-species Lotka-Volterra
  model by means of Monte Carlo simulations on a two-dimensional
  lattice.  Individuals are assigned inheritable predation efficiencies; 
  quenched randomness in the spatially varying reaction rates serves 
  as environmental noise.  We find that environmental variability 
  enhances the population densities of both predators and prey 
  while demographic variability leads to essentially neutral optimization.
\end{abstract}

\maketitle

The mathematical modeling of species interactions continues to be a
central issue in population 
ecology~\cite{Murray2002,Hofbauer1998,Smith1974,may2001stability}.
Several simple models have been proposed, investigated, and sometimes
realized under laboratory conditions. Yet more realistic and thus
biologically more relevant model variants obviously have to include both 
external spatial disorder in the reaction rates to account for varying 
environmental conditions and intrinsic demographic heterogeneity 
stemming from trait variability in individuals. While
we addressed the former in a recent study~\cite{Dobramysl2008}, our
goal in this letter is to investigate the interplay between quenched
spatial rate disorder and {\em additional} variability of individuals'
reaction rates, as well as intriguing evolutionary co-optimization
within interacting populations. 

We focus on the Lotka-Volterra (LV) predator-prey model owing to its
simplicity and because its basic features are well-understood. It was
first introduced to study fish populations in the Adriatic sea and chemical
oscillations~\cite{Lotka1920,Volterra1926}. While the original deterministic
LV (mean-field) equations yield neutral cycles and hence persistent 
nonlinear oscillations around a marginal fixed point~\cite{Murray2002}, in
stochastic implementations this species coexistence fixed point becomes 
stable and is approached very slowly through damped 
oscillations~\cite{Provata1999, Rozenfeld1999, Lipowski1999,Droz2001,
  Antal2001, Kowalik2002, McKane2005}. Spatially extended stochastic 
versions of the LV model yield striking dynamical patterns and emergent 
inter-species correlations~\cite{Matsuda1992,Satulovsky1994, 
  Boccara1994,Durrett1999,Mobilia2006,Mobilia2006a,Washenberger2007} 
which may be utilized to quantitatively assess the response to external or 
internal changes. Population stability can be measured via the extinction 
time in small systems, where the stochastic kinetics ultimately reaches an 
absorbing zero-particle state~\cite{Mobilia2006a,Dobrinevski2012}.

In our study of the effects of environmental rate variability in the LV model, 
we found a remarkable increase of the asymptotic population densities of 
\emph{both} species with enhanced quenched spatial disorder, i.e., 
predation rates that are fixed to different lattice sites~\cite{Dobramysl2008}. 
Yet the observed erratic population oscillations and relaxation towards the
(quasi-)steady state occur on the time scale of many generations; for real
biological systems, one therefore needs to address Darwinian evolutionary 
adaptation of individuals' traits. Consequently, we introduce fundamentally 
novel features by endowing \emph{individual} predator and prey particles 
with randomly selected rates, and investigate whether and how optimization 
within each species due to imperfect efficiency inheritance (mimicking random
mutations) further reinforces the total population's stability and fitness. 
Dynamical coevolution of interacting species is a crucial feature of adapting 
ecological systems and has been studied 
experimentally~\cite{Yoshida2003,Kishida2006} as well as
theoretically~\cite{Weitz2005,Rogers2012,Traulsen2012,Fort2012}. 
Combining quenched spatial with individual, evolving rate distributions allows 
us to quantitatively assess the relative importance of environmental vs. 
demographic, inheritable variabilities in a nonlinear competing two-species
predator-prey system.

We find that {\em both} environmental and individual-based
variabilities combined with random mutations produce a marked
enhancement of the quasi-stationary densities of both species, thus
considerably extending our earlier conclusions for purely
environmental randomness~\cite{Dobramysl2008}. In addition, individual
variability stabilizes both predator and populations against
extinction. Remarkably, the optimization of predation and evasion
capabilities of either species turns out to be essentially neutral in
the population densities; in contrast to genetic drift
models~\cite{Hartl1997}, our nonlinear model does not lead to trait
fixation.

We consider a spatially extended version of the LV model consisting of two
particles species.  The ``predator'' species is subject to spontaneous decay
$A\rightarrow\emptyset$ with rate $\mu$, while the ``prey'' B reproduce 
(asexually): $B\rightarrow2B$ with rate $\sigma$. Different particles interact
on-site with a non-uniform predation rate $\lambda$, whereupon a prey is 
removed and replaced with a predator: $A+B\rightarrow2A$. The prey birth
and predator death rates both remain fixed at a uniform value
$\sigma=\mu=0.5$ for all particles and lattice sites, whereas the predation 
rates are allowed to vary between different positions and participating
particles (see below). Particles exist on a two-dimensional square lattice with
128 $\times$ 128 sites and periodic boundary conditions. (We could not find 
significant finite-size effects already at this lattice size.)  Both species
perform unbiased random walks via nearest-neighbor hopping occurring with
probability one, hence all rates are to be understood relative to the
diffusivity $D$.  Reactions occur on-site, assuming infinite local carrying 
capacities, implying that the growth of the population on any single site is
essentially unrestricted (with a safety limit of $n_i\le1000$ per lattice site 
$i$ that is never reached with the parameters investigated in the present
study). The predator extinction transition occurring in model variants with
restricted site occupation is thus absent 
here~\cite{Antal2001,Mobilia2006a,Washenberger2007}. The initial 
population distribution of both predator and prey particles is chosen randomly
with a mean density $\rho_{A,i}=\rho_{B,i}=1$. The simulation proceeds 
via random sequential updates, with one Monte Carlo step being completed 
when on average each particle in the simulation has moved and had a 
chance to react~\footnote{More technical details about the algorithm can 
be found in Ref.~\cite{Washenberger2007}.}.

In order to model variability of individuals and trait inheritance, each 
particle carries a predation efficacy property $\eta\in[0,1]$, 
determined during the particle's creation and providing a coarse-grained
characterization of the combined efficacies of its genetic heritage 
(genes) and its learned strategies (memes).  An offspring's position in
efficiency space will thus be near its parent's location but subject to 
random changes (mutations in the case of genes, adaptations of 
strategies in the case of memes), thereby suggesting the use of a
normalized Gaussian distribution centered at the parent's efficiency 
value $\eta_P$ (truncated to the interval $[0,1]$ accessible to a 
reaction probability) to assign an efficiency value $\eta_O$ to the 
offspring during reproduction.  The standard deviation $w_P$ of the 
Gaussian function constitutes a model parameter and corresponds to the 
average severity of mutations from one generation to the next.  Note 
that the efficiency assigned to a particle $\eta$ is different from the 
traditional genetic fitness, which is defined as the average number of 
offspring produced by a genome.  It represents a mesoscopic 
continuous stochastic variable, as opposed to a genetic description 
employing naturally discrete values.

Since we wish to address the distinctions between internal and spatial
randomness, we introduce in addition environmental variability by
assigning a spatial predation efficacy value $\eta_S$ to each lattice
site, drawn from a normalized Gaussian distribution with fixed mean 
0.5 and standard deviation $w_S$, truncated to $[0,1]$, and set to 
be fixed in time~\footnote{For more details on the implementation of 
spatial variability see Ref.~\cite{Dobramysl2008}}. The ensuing 
predation rate $\lambda$ is a random variable as well, namely a 
function of both the spatial efficiency at the lattice site the reaction 
occurs on and the two individual predation efficacies of the participating 
predator and prey particles.  We finally define a model parameter 
$\zeta$ that describes the relative importance of the spatial over
individual efficacies:
\begin{equation}
  \label{eq:1}
  \lambda = \zeta \, \eta_S + (1-\zeta) \, (\eta_A+\eta_B)/2 \, .
\end{equation}

\begin{figure}[tb]
  \centering
  \includegraphics[width=\columnwidth]{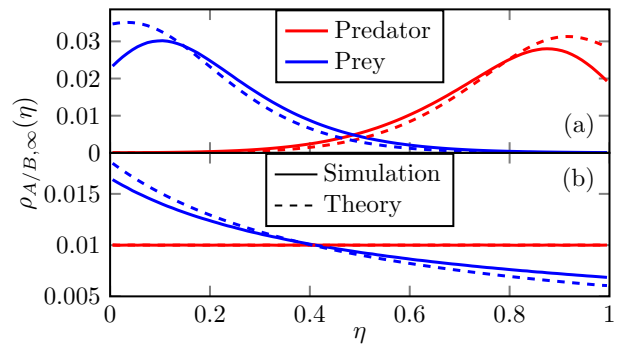}
  \caption{(Color online.) (a) The predator (red) and prey (blue)
    population densities in the (quasi-)steady state as functions of the
    predation efficiency, averaged over $10^4$ Monte Carlo simulation
    runs, optimize towards high $\overline{\eta}_A\approx0.71$ and low
    $\overline{\eta}_B\approx0.26$ mean values, respectively.  The
    standard deviation of the Gaussian inheritance distribution is
    $w_P=0.1$ and the spatial influence factor $\zeta=0$.  In contrast
    to genetic drift models, both densities do not fixate at extreme 
    predation efficiencies ($\eta = 0,1$). The dashed lines show the 
    theoretical prediction from an effective stochastic mean-field model.  
    (b) For a flat inheritance distribution ($w_P=\infty$), the predators 
    experience no selection bias while the prey population is 
    preferentially selected towards low predation efficacies.}
  \label{fig:popdistrib}
\end{figure}
Over many generations, species optimize their predation / evasion
efficiency through evolving their efficacy distributions by means of
random inheritance.  Figure~\ref{fig:popdistrib}(a) shows the
population density histograms $\rho_{A/B}(\eta)$ for a representative
case with moderate inheritance variability $w_P=0.1$.  The initial
population of predator and prey particles had an assigned predation
efficacy of $\eta_{A/B}=0.5$.  We have carefully checked that the 
final (quasi-)steady state population distribution does not depend on 
the (uncorrelated) initial conditions (except for those rare simulation 
runs when either the prey or predator population went extinct).  The 
predator population maximum moves towards higher mean efficacy 
whereas the prey population, for which lower values of the predation 
efficiency are preferable, tends towards a lower average.  Predators 
with a slightly higher efficiency value are more successful at predation 
and thereby reproduce more often.  Hence their improved predation 
capability is passed on to subsequent generations with a higher 
frequency, driving the overall predator population toward higher 
efficiency values. Similarly, prey particles with a lower predation 
efficiency are better at evasion and thus survive longer.  This gives
them the chance to reproduce at a higher rate, driving the prey 
population towards low mean efficacy.
In the extreme situation of completely random assignment of predation
efficiencies, where no correlations between the corresponding values 
for parents and offspring are implemented (equivalent to a uniform 
inheritance distribution with $w_P=\infty$), we already see a strong 
tendency towards low efficacies for the prey species; see 
Fig.~\ref{fig:popdistrib}(b).  This feature is explained by the bias in 
the predation rule that favors selection of prey particles with higher 
efficiency values. For predators no such bias exists; hence their 
population distribution in efficiency space remains flat. Spatial 
fluctuations modify the results quantitatively, but not qualitatively, 
whence we observe a slightly more pronounced effect in the prey 
distribution in our studies of non-spatial systems.

\begin{figure}[tb]
  \centering
  \includegraphics[width=\columnwidth]{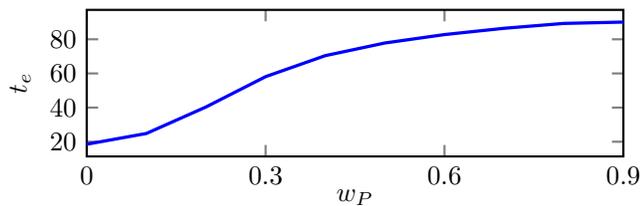}
  \caption{(Color online.) Mean extinction time $t_e$ of a small system 
    of $10 \times 10$ lattice sites as a function of individual variability 
    $w_P$ and for $\zeta=0$.  Extinction here is defined as the 
    event when either species goes extinct.  With increasing variability 
    the system on average takes longer to go extinct.}
  \label{fig:extinction}
\end{figure}
In order to analytically verify our data we write down the mean-field
equations for a well-mixed zero-dimensional LV system with individual
variability.  We consider the number of particles of species A and B as a 
function of predation efficiency.  The predation efficiency $\eta$ is a 
bounded quantity in the interval $[0,1]$ assigned to particles at their 
time of creation.  We discretize this interval into $N$ bins with a spacing 
of $\Delta\eta=1/N$ and midpoint $\eta_i=(i+1/2)/N$, and denote the 
number of A and B particles in a bin $i$ respectively as $a_i$ and $b_i$ 
($i = 0,\ldots,N-1$).  To model individual variability we introduce the
probability $f_{ij}=f(\eta_i,\eta_j)$ for a parent with efficiency 
$\eta_i$ to produce offspring with efficiency $\eta_j$.  The predation
rate is a function of the efficacies of the predator A and prey B 
participating in the predation reaction: 
$\lambda_{ij} = (\eta_i+\eta_j)/2$ [essentially the discretized 
Eq.~(\ref{eq:1}) with $\zeta=0$].  Thus we arrive at the coupled
mean-field rate equations for the case of purely individual variability:
\begin{eqnarray}
  \label{eq:meanfielda}
  \dot{a}_i&=-\mu a_i+\sum_j\sum_k\lambda_{jk}f_{ik}a_kb_j \, , \\
  \label{eq:meanfieldb}
  \dot{b}_i&=\sigma\sum_kf_{ik}b_k-\sum_j\lambda_{ij}a_jb_i \, .
\end{eqnarray}
The steady-state densities are obtained by setting the time
derivatives to zero, yielding expressions for $a_i$ and $b_i$ that can
be solved iteratively for any inheritance probability distribution $f$,
as shown (dashed) in Fig.~\ref{fig:popdistrib}(a).

In the special case of a uniform probability distribution (implying the 
absence of any correlation between the predation efficiencies of parent 
and offspring particles) $f_{ij}=1/N$, the steady-state densities, 
defined by $\rho_{A,i}=a_i/\sum_ja_j$ and $\rho_{B,i}=b_i/\sum_jb_j$, 
can be obtained exactly.  The predator population acquires a constant 
($\eta$-independent) value of $\rho_{A}=1/N$, whereas the prey 
population decreases with increasing $\eta$ as 
$\rho_{B}(\eta)=\frac{2}{N\ln3}\frac{1}{1+2\eta}$.  This result is 
exactly mirrored by our zero-dimensional Monte Carlo simulations.  In 
spatially extended systems fluctuations modify the density distributions, 
leading to a prey density dependence that is slightly less steep as a
function of the predation efficiency, see Fig.~\ref{fig:popdistrib}(b).  
Correlation effects not captured by mean-field theory are evidently 
strongest at the distribution maxima.

We collected extinction time histograms for small systems (lattice size 
10 $\times$ 10 sites) to determine the influence of individual variability 
on the stability of the population.  In finite stochastic systems with an 
absorbing state (here, predator extinction), fluctuations will eventually 
drive the system into the absorbing state.  Figure~\ref{fig:extinction} 
demonstrates that the mean extinction time is enhanced by a factor up 
to $\approx 4.5$ by individual variability, rendering the system 
markedly more robust against extinction.

\begin{figure}
  \includegraphics[width=\columnwidth]{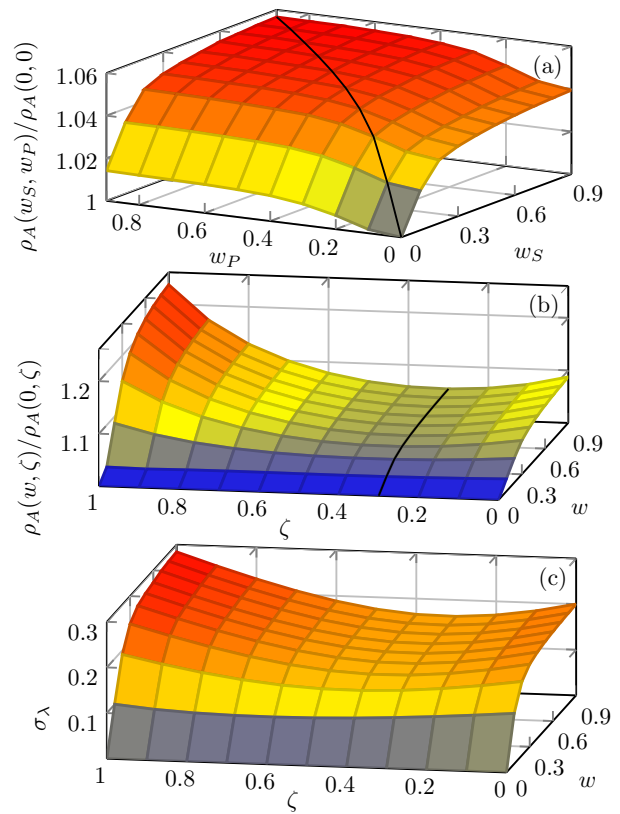}
  \caption{(Color online.) (a) The (quasi-)steady state predator 
    density $\rho_{A}$ as a function of the spatial and individual 
    variability $w_S$ and $w_P$ for $\zeta=0.3$. The black line 
    represents the slice of equal spatial and individual variability 
    $w_S=w_P$ from the minimum in (b). (b) The predator density 
    shows a consistent increase for all values of the spatial variability 
    influence $\zeta$ as function of equal variabilities $w=w_S=w_P$ 
    over a system with zero variability.  A remarkable minimum is 
    observed near $\zeta=0.3$ (black line).  (c) The standard 
    deviation of the predation rate $\sigma_\lambda$, calculated via 
    error propagation from the spatial and individual predation 
    efficiency distributions.}
  \label{fig:densitysigma}
\end{figure}
To quantify the influence of variability, and in particular the distinction
between individual (internal) variability and spatial environmental 
randomness, we measured its impact on the (quasi-)steady state 
particle density for both species. Figure~\ref{fig:densitysigma}(a) 
displays the relative change of the predator density 
$\rho_{A}(w_S,w_P)$ over the zero-variability case as a function of 
$w_S$ and $w_P$ for $\zeta=0.3$. Both types of variability contribute 
additively and positively to the density enhancement. 
Figure~\ref{fig:densitysigma}(b) shows the relative density change as
a function of $w=w_S=w_P$ and $\zeta$~\footnote{See 
  Supplementary Material at URL for a spatial animation of a 
  representative realization.  The insert in the movie shows the 
  temporal evolution of the population distributions.}. The prey 
density shows the same quantitative behavior for all parameter ranges.
Hence we observe a significant increase of the population densities of
\emph{both} species for higher variability not only for purely spatial
($\zeta=1$) randomness~\cite{Dobramysl2008}, but also for individual
variability ($\zeta=0$). In contrast, the effect of spatial randomness in 
either the prey birth rate $\sigma$ or the predator death rate $\mu$ 
on the species densities stayed below a rather low value of $2\%$.

The striking minimum in the density increase occurring near a spatial
influence factor $\zeta=0.3$ arises from the combined variabilities
through the quenched randomness of the lattice and the emergent
variability of the individual particles.  We argue that the density
increase is primarily a monotonic function of the variability in the
predation rate $\lambda$. Using the dependence of the predation rate
$\lambda$ on the spatial predation efficacy value $\eta_S$ and the
predation efficiencies of the participating particles $\eta_A$ and
$\eta_B$ given in the text, the standard deviation of $\lambda$ is 
$\sigma_\lambda=\sqrt{\zeta^2 \, \sigma_S^2 + (1-\zeta)^2 \, 
  (\sigma_A^2 + \sigma_B^2) / 2}$.
Due to the truncation of predation efficiency values to the range
$[0,1]$, the effective standard deviation of the spatial predation
efficacy is different from the environmental variability parameter. 
Similarly, the standard deviation of the predation efficiencies of 
individual particles have to be taken from simulation data. 
Figure~\ref{fig:densitysigma}(c) shows the resulting standard 
deviation of $\lambda$ as a function of $w$ and $\zeta$ which 
is a measure of the effective combined variability. It reflects the
minimum in the density increase at $\zeta\approx 0.3$.  The data 
also emphasize that environmental variability has a more 
pronounced effect on the species densities as compared to 
demographic variability, since the density increase is 
disproportionally higher for $\zeta\rightarrow 1$.

Surprisingly, we observe that low individual variability with weak or
no spatial influence, i.e. $0<w_P\ll1$ and $\zeta=0$, yields the
strongest species optimization with the maxima of the predator and
prey populations closest to $\eta=1$ and $\eta=0$ respectively; see
Fig.~\ref{fig:popdistrib}(a). But the enhancement of the overall
species densities in this regime is minute and tends to zero for small
$w_P$; see the lower right corner of Fig.~\ref{fig:densitysigma}(b). 
The respective benefits of the up- / downward optimization of the
predator / populations in terms of predation efficiency clearly almost
cancel each other. Hence we conclude that predation efficiency 
optimization is essentially neutral and carries no benefit for either 
species in terms of their net population densities (at least in the 
context of our model), despite its vital necessity to ensure the 
survival of coevolving species. This also reinforces our argument 
that the density enhancement is a function of rate variability only.

\begin{figure}[tb]
  \centering
  \includegraphics[width=\columnwidth]{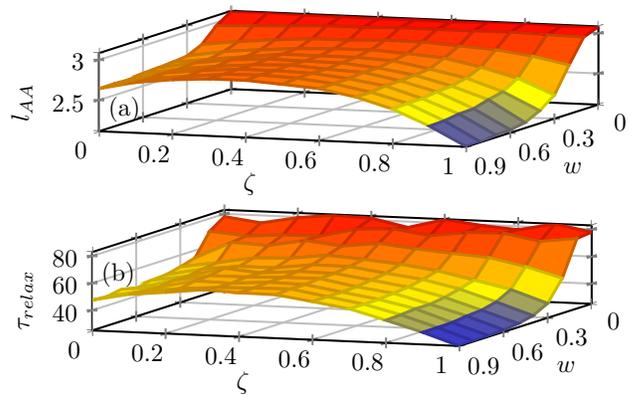}
  \caption{(Color online.) (a) The correlation length $l_{AA}$ from the 
    predator-predator autocorrelation function.  The prey-prey 
    correlation length $l_{BB}$ would essentially display the same 
    shape as $l_{AA}$ scaled by $\sim0.9$.  
    (b) The predator density relaxation time $\tau_{relax}$ toward 
    the quasi-stationary state.}
  \label{fig:corrlengths}
\end{figure}
The predator-predator and prey-prey correlation lengths $l_{AA}$ 
and $l_{BB}$ and typical predator-prey distance $l_{AB}$, measured 
by extracting the (quasi-)steady state exponential decay length and 
the position of the maximum (in the case of $l_{AB}$) from the 
correlation functions $C_{\alpha\beta}(x)=\left<n_{\alpha i+x}
n_{\beta i}\right> - \rho_\alpha \rho_\beta$ with 
$\alpha,\beta=A,B$~\footnote{Figs.~4-6 of Ref.~\cite{Mobilia2006a}
  and Fig.~5 in Ref.~\cite{Washenberger2007} show these correlation
  functions for LV systems without disorder.}, decrease for increasing 
variability $w$; see Fig.~\ref{fig:corrlengths}(a).  For 
$\zeta=1$ we reproduce the data from Ref.~\cite{Dobramysl2008}, 
where we argued that the decrease in $l_{\alpha\beta}$ indicated a
more tightly clustered population around lattice sites with small spatial 
predation efficiency $\eta_S$, leading to the observed enhanced 
densities and higher amplitudes in the initial population oscillations. 
Surprisingly we also see a (less pronounced) decrease of 
$l_{\alpha\beta}$ for $\zeta\rightarrow0$, indicating the existence of 
spontaneously formed tight activity patches around clusters of highly 
optimized prey particles.  To investigate the effect of the combined 
variability on the relaxation of the population densities we collected 
data on the characteristic decay time of the initial predator population 
oscillations by least-square fitting of an exponentially decaying
sinosiodal function to the predator species density time series; see 
Fig.~\ref{fig:corrlengths}(b).  As expected, increasing disorder 
$w$ induces a roughly threefold decrease in the purely spatial 
case ($\zeta=1$) and about a twofold decrease in the individual 
variability case ($\zeta=0$).

In conclusion we performed an extensive numerical Monte Carlo
simulation study to assess how external environmental randomness and
individual variability modified through mutations during inheritance
compete and affect the coevolutionary population dynamics of two 
coexisting species in a spatial stochastic LV model. The overall predator 
and prey densities are {\em both} enhanced by environmental 
variations, while evolutionary optimization within each species has an 
essentially neutral net effect.  To better understand this evolutionary
trait optimization, we derived a mean-field model that qualitatively 
reproduces our simulation results.  In addition we find that increased 
individual variability stabilizes both populations against extinction.  
There are certainly other intriguing aspects pertaining to 
variability in ecological models that deserve further investigation, 
promising amazingly rich features and crucial quantitative insight.

We gratefully acknowledge inspiring discussions with G. Daquila,
E. Frey, J. Phillips, T. Platini, M. Pleimling, B. Schmittmann, and
R.K.P. Zia.


\begin{thebibliography}{29}

\bibitem{Murray2002}
J.~D. Murray, \emph{{Mathematical Biology}}, 3rd ed., Vol. I and II
  (Springer, New York, 2002).

\bibitem{Hofbauer1998}
J.~Hofbauer and K.~Sigmund, \emph{{Evolutionary Games and Population Dynamics}}
  (Cambridge University Press, Cambridge, 1998).

\bibitem{Smith1974}
J.~Smith, \emph{{Models in ecology}} (Cambridge University
  Press, Cambridge, 1974) p. 145.

\bibitem{may2001stability}
R.~May, \emph{{Stability and complexity in model ecosystems}}, Vol.~6
  (Princeton University Press, Princeton, NJ, 1973).

\bibitem{Dobramysl2008}
U.~Dobramysl and U.~C. T\"{a}uber, Phys. Rev. Lett. \textbf{101}, 258102
  (2008).

\bibitem{Lotka1920}
A.~J. Lotka, J. Am. Chem. Soc. \textbf{42}, 1595 (1920).

\bibitem{Volterra1926}
V.~Volterra, Mem. Accad. Sci. Lincei. \textbf{2}, 31 (1926).

\bibitem{Provata1999}
A.~Provata, G.~Nicolis, and F.~Baras, J. Chem. Phys. \textbf{110},
  8361 (1999).

\bibitem{Rozenfeld1999}
A.~Rozenfeld and E.~Albano, Physica A \textbf{266}, 322 (1999).

\bibitem{Lipowski1999}
A.~Lipowski, Phys. Rev. E \textbf{60}, 5179 (1999).

\bibitem{Droz2001}
M.~Droz and A.~Pekalski, Phys. Rev. E \textbf{63}, 051909 (2001).

\bibitem{Antal2001}
T.~Antal and M.~Droz, Phys. Rev. E \textbf{63}, 056119 (2001).

\bibitem{Kowalik2002}
M.~Kowalik, A.~Lipowski, and A.~L. Ferreira, Phys. Rev. E \textbf{66},
  066107 (2002).

\bibitem{McKane2005}
A.~J. McKane and T.~J. Newman, Phys. Rev. Lett. \textbf{94}, 218102
  (2005).

\bibitem{Matsuda1992}
H.~Matsuda, N.~Ogita, A.~Sasaki, and K.~Sat\=o, Progr. Theoret. Phys.
  \textbf{88}, 1035 (1992).

\bibitem{Satulovsky1994}
J.~E. Satulovsky and T.~Tom\'{e}, Phys. Rev. E \textbf{49}, 5073 (1994).

\bibitem{Boccara1994}
N.~Boccara, O.~Roblin, and M.~Roger, Phys. Rev. E \textbf{50}, 4531
  (1994).

\bibitem{Durrett1999}
R.~Durrett, SIAM Review \textbf{41}, 677 (1999).

\bibitem{Mobilia2006}
M.~Mobilia, I.~T. Georgiev, and U.~C. T\"{a}uber, Phys. Rev. E
  \textbf{73}, 040903(R) (2006{\natexlab{a}}).

\bibitem{Mobilia2006a}
M.~Mobilia, I.~T. Georgiev, and U.~C. T\"{a}uber, J. Stat.
  Phys. \textbf{128}, 447 (2006{\natexlab{b}}).

\bibitem{Washenberger2007}
M.~J. Washenberger, M.~Mobilia, and U.~C. T\"{a}uber, J. Phys.
  Cond. Matter \textbf{19}, 065139 (2007).

\bibitem{Dobrinevski2012}
A.~Dobrinevski and E.~Frey, Phys. Rev. E \textbf{85}, 051903 (2012).

\bibitem{Yoshida2003}
T.~Yoshida, L.~E. Jones, S.~P. Ellner, G.~F. Fussmann, and N.~G. Hairston,
  Nature \textbf{424}, 303 (2003).

\bibitem{Kishida2006}
O.~Kishida, Y.~Mizuta, and K.~Nishimura, Ecology \textbf{87}, 1599 (2006).

\bibitem{Weitz2005}
J.~S. Weitz, H.~Hartman, and S.~A. Levin, Proc. Natl. Acad. Sci. USA \textbf{102}, 9535 (2005).

\bibitem{Rogers2012}
T.~Rogers, A.~J. McKane, and A.~G. Rossberg, Europhys. Lett. \textbf{97},
  40008 (2012).

\bibitem{Traulsen2012}
A.~Traulsen, J.~C. Claussen, and C.~Hauert, Phys. Rev. E \textbf{85},
  041901 (2012).

\bibitem{Fort2012}
H.~Fort and P.~Inchausti, J. Stat. Mech. Theor. Exp. \textbf{2012}, P02013 (2012).

\bibitem{Hartl1997}
D.~L. Hartl and A.~G. Clark, \emph{Principles of Population
  Genetics.}, 3rd ed.
  (Sinauer Associates, Sunderland, MA, 1997) Chap. 274-275.

\end{thebibliography}
\end{document}